\documentclass[showpacs,preprint,amsmath,amssymb,aps,11pt]{revtex4}

\usepackage{graphicx}
\usepackage{graphicx,rotating,lscape}

\begin{document}

\def\erf{\mathrm{erf}}

\title{An analytic parametrization of the hyperonic matter equation of state}

\author{Isaac Vida\~na}

\affiliation{Centro de F\'{i}sica Computacional, Department of Physics, University of Coimbra, PT-3004-516 Coimbra, Portugal}

\author{Domenico Logoteta}

\affiliation{Centro de F\'{i}sica Computacional, Department of Physics, University of Coimbra, PT-3004-516 Coimbra, Portugal}

\author{Constan\c{c}a Provid\^{e}ncia}

\affiliation{Centro de F\'{i}sica Computacional, Department of Physics, University of Coimbra, PT-3004-516 Coimbra, Portugal}

\author{Artur Polls}
\affiliation{Departament d'Estructura i Constituents de la Mat\`eria and Institut de Ci\`{e}ncies del Cosmos, Universitat de Barcelona, Avda. Diagonal 647, E-08028 Barcelona, Spain}

\author{Ignazio Bombaci}
\affiliation{Dipartimento di Fisica ``E. Fermi'', Universit\`a di Pisa, and INFN, Sezione di Pisa, Largo B. Pontecorvo 3, I-56127 Pisa, Italy}

\begin{abstract}

An analytic parametrization of the hyperonic matter equation of state based on microscopic Brueckner--Hartree--Fock calculations has been constructed using the realistic
Argonne V18 nucleon-nucleon potential plus a three-body force of Urbana type, and three models of the hyperon-nucleon interaction: the Nijmegen soft-core models NSC89 and 
NSC97e, and the most recent meson-exchange potential of the J\"{u}lich group. The construction of this 
parametrization is based on a simple phase-space analysis and reproduces with good accuracy the results of the microscopic calculations with a small number of parameters. 
This parametrization allows for rapid calculations that accurately mimic the microscopic results, being therefore, very useful from a practical point of view.

\end{abstract}

\vspace{0.5cm}

\maketitle


Neutron stars offer an interesting interplay between nuclear processes and astrophysical observables \cite{shapiro83,prakash97}. 
Properties of neutron stars, such as the mass range, the mass-radius relationship, the moment of inertia,
the crust thickness or the cooling rate, are closely related to the underlying nuclear matter equation of 
state (EoS) for a wide range of densities and temperatures \cite{chamel08}. Thus, its determination is an essential ingredient
for understanding such properties. 

At densities near the saturation density of nuclear matter, neutron star matter is thought to be mainly composed of neutrons, 
protons and leptons (electrons and muons) in $\beta$-equilibrium. As density increases, new hadronic degrees of freedom may appear 
in addition to nucleons. Hyperons, baryons with a strangeness content, are an example of these degrees of freedom. Contrary to 
terrestial conditions, where hyperons are unstable and decay into
nucleons through the weak interaction, the equilibrium conditions in neutron stars can make the inverse
process, {\it i.e.,} the conversion of nucleons into hyperons, happen, so the formation of hyperons becomes energetically favorable. 
Although hyperonic matter is an idealized physical system, the theoretical determination of the corresponding EoS is an essential step 
towards the understanding of  properties of neutron stars. Moreover, the comparison of theoretical predictions for 
the properties of these objects with the observations can provide strong constraints on the interactions among 
their constituents. Since the pionneering work of Ambartsumyan and Saakyan \cite{AS60} the EoS of hyperonic
matter has been considered by several authors either from phenomenological 
\cite{glendenning,weber89,glendenning91,knorren95,schaffner96,huber98,balberg97,balberg99}
or microscopic \cite{schulze95,baldo00,vidana00,vidana00b,sammarruca09,dapo10} approaches. 

In phenomenological approaches the input is a density-dependent interaction which contains a certain
number of parameters adjusted to reproduce experimental data. Within this approach Balberg and Gal \cite{balberg97}
derived an analytic effective EoS using density-dependent baryon-baryon potentials based on Skyrme-type
forces including hyperonic degrees of freedom. The features of this EoS rely on the properties of nuclei
for the nucleon-nucleon (NN) interaction, and mainly on the experimental data from hypernuclei for the
hyperon-nucleon (YN) and hyperon-hyperon (YY) interactions. This EoS reproduces characteristic properties of high-density
matter found in theoretical microscopic models. Within the same scheme, several authors \cite{lanskoy97,cugnon00} have developed
Skyrme-like YN potentials to study properties of single- and multi-$\Lambda$ hypernuclei with the Skyrme--Hartree--Fock formalism.

An alternative phenomenological approach involves the formulation of an effective relativistic mean field 
theory (RMFT) of interacting hadrons \cite{serot}. This  fully relativistic approach treats the baryonic and mesonic degrees of
freedom explicitely, and is, in general, easier to handle because it only involves local
densities and fields. The EoS of dense matter with hyperons was first described within the RMFT by Glendenning
\cite{glendenning} and latter by other authors \cite{weber89,glendenning91,knorren95,schaffner96,huber98}. The parameters in this approach are fixed by the properties of 
nuclei and nuclear bulk matter for the nucleonic sector, whereas the coupling constants of the hyperons are fixed by 
symmetry relations, hypernuclear observables and compact star properties. 

In microscopic approaches, on the other hand, the input are two-body baryon-baryon interactions that describe the
scattering observables in free space. These realistic interactions have been mainly constructed  within the framework
of a meson-exchange theory, although recently a new approach based on chiral perturbation theory has emerged as
a powerful tool. In order to obtain the EoS one has to solve the complicated nuclear many-body problem \cite{baldo99,muether00}. A great
difficulty of this problem lies in the treatment of the repulsive core, which dominates the short-range
behavior of the interaction. Various methods have been considered to solve the nuclear many-body problem: 
the variational approach \cite{akmal98}, the correlated basis function (CBF) formalism \cite{fantoni93}, 
the self-consistent Green's function (SCGF) technique \cite{kadanoff62}, or the Brueckner--Bethe--Goldstone (BBG) \cite{brueckner68} and the 
Dirac--Brueckner--Hartree-Fock (DBHF) theories \cite{haar87}. Nevertheless, although all of them have been extensively applied to the 
study of nuclear matter, up to our knowledge, only the BBG theory \cite{schulze95,baldo00,vidana00,vidana00b}, and very recently 
the DBHF one \cite{sammarruca09}, and the $V_{low k}$ approach \cite{dapo10}, have been extended to the hyperonic sector. 

The microscopic approach is in general technically complex and very time consuming. Therefore, from a practical
point of view, it would be interesting to have an analytic parametrization of the hyperonic matter EoS based on such approach 
that allow to mimic the microscopic results in a fast way with a small number of parameters. In the present work we will build 
a density functional for the EoS based on microscopic Brueckner--Hartree--Fock (BHF) calculations of hyperonic matter. In addition 
to the nucleonic degress of freedom we will consider only $\Lambda$ and $\Sigma^-$ hyperons in the construction of our functional, 
the reason being that these two types of hyperons are the ones appearing first in calculations of $\beta$-stable neutron 
star matter based on microscopic approaches \cite{schulze95, baldo00, vidana00,vidana00b}. The other hyperons, $\Sigma^0$,
$\Sigma^+$, $\Xi^0$ and $\Xi^-$, being heavier, either do not appear or only show up at very large densities in microscopic calculations.

After a brief review of the BHF approach to the EoS, we will detail the construction of the functional. 
We will finish by testing and discussing the quality and validity of our parametrization.


Our calculations are based on the BHF approximation of the BBG theory extended to the hyperonic matter case
\cite{schulze95, baldo00, vidana00,vidana00b}. Therefore, our-many body scheme starts by constructing all the 
baryon-baryon $G$ matrices, which describe in an effective way the interaction between two baryons in the presence of 
a surrounding medium. The $G$ matrices can be obtained by solving the Bethe--Goldstone equation, written 
schematically as
\begin{equation}
G(\omega)_{B_1B_2,B_3B_4}=V_{B_1B_2,B_3B_4}+\sum_{B_iB_j}V_{B_1B_2,B_iB_j}
\frac{Q_{B_iB_j}}{\omega-E_{B_i}-E_{B_j}+i\eta}G(\omega)_{B_iB_j,B_3B_4} \ ,
\label{eq:bge}
\end{equation}
where the first (last) two subindices indicate the initial (final) two-baryon states compatible with a given value
$S$ of the strangeness, namely NN for $S=0$ and YN for $S=-1$; $V$ is the bare baryon-baryon 
interaction (NN or YN); $Q_{B_iB_j}$ is the Pauli operator, that prevents the intermediate
baryons $B_i$ and $B_j$ from being scattered to states below their respective Fermi momenta; and the starting energy
$\omega$ corresponds to the sum of the nonrelativistic single-particle energies of the interacting baryons. 
We note here that, although we have considered only $\Lambda$ and $\Sigma^-$ hyperons in the construction of our parametrization of
the EoS, $\Sigma^0$ and $\Sigma^+$ hyperons have been also taken into account in the intermediate YN states when solving the 
Bethe--Goldstone equation. The interested reader is referred to Refs.\ \cite{schulze95, baldo00, vidana00,vidana00b} for computational 
details.

The single-particle energy of a baryon $B_i$ is given by 
\begin{equation}
E_{B_i}(\vec{k})=M_{B_i}+\frac{\hbar^2k^2}{2M_{B_i}}+U_{B_i}(\vec{k})\ .
\label{eq:spe}
\end{equation}
Here $M_{B_i}$ denotes the rest mass of the baryon, and the single-particle potential $U_{B_i}$ represents
the averaged field ``felt'' by the baryon owing to its interaction with the other baryons of the medium. In the BHF approximation,
$U_{B_i}$ is given by  
\begin{equation}
U_{B_i}(\vec{k})=\sum_{B_j}U_{B_i}^{(B_j)}(\vec{k})=\mbox{Re}\,\sum_{B_j}\sum_{\vec{k'}}n_{B_j}(|\vec{k'}|)
\langle \vec{k}\vec{k'}|G(\omega)_{B_iB_j,B_iB_j}(\omega=E_{B_i}(\vec{k})+E_{B_j}(\vec{k'}))|\vec{k}\vec{k'} \rangle \ ,
\label{eq:spp}
\end{equation}
where a sum over all the different partial contributions, $U_{B_i}^{(B_j)}(\vec{k})$, is performed, $n_{B_j}(|\vec{k}|)$ is the occupation number of the species $B_j$, 
and the matrix elements are properly antisymmetrized when baryons $B_i$ and $B_j$ belong to the same isomultiplet. We note here that the so-called continuous
prescription has been adopted for the single-particle potentials when solving the Bethe--Goldstone equation, since, as shown by the authors
of Refs.\ \cite{song98,baldo00b}, the contribution to the energy per particle from three-body clusters is minimized in this prescription. 

All the calculations carried out in this work have been performed with the realistic Argonne V18 \cite{wiringa95} NN interaction supplemented with a NNN 
three-body force (TBF) of Urbana type which, for use in BHF calculations, was reduced to a two-body density-dependent force by averaging over the third nucleon 
in the medium \cite{baldo99b}. This TBF contains two parameters that are fixed by requiring that the BHF calculation reproduces the energy and saturation density 
of symmetric nuclear matter. The interested reader is referred to the works of Refs. \cite{zhou04,li08,li08b} for a recent analysis of the use of TBF's in 
nuclear and neutron matter. The YN $G$-matrices have been constructed using three YN models: the Nijmegen soft-core models NSC89 \cite{maessen89} and
NSC97e \cite{stoks99}, and the most recent YN meson-exchange potential of the J\"{u}lich group \cite{haidenbauer05}. In the following we will use the
names J\"{u}lich, NSC89 and NSC97 models to denote the three different NN+NNN+YN models under consideration, since the pure nucleonic part is the same in 
all of them. We note that the YY interaction has not been considered in the present work due to the large uncertainties still existing in this sector. 

Once a self-consistent solution of Eqs.\ (\ref{eq:bge}) and (\ref{eq:spp}) is obtained, the total energy per particle is easily calculated:
\begin{equation}
\frac{E}{A}=\frac{1}{A}\sum_{B_i}\sum_{\vec{k}}n_{B_i}(|\vec{k}|)\left[
\frac{\hbar^2k^2}{2M_{B_i}}+\frac{1}{2}U_{B_i}(\vec{k})\right] \equiv \frac{T}{A}+\frac{V}{A} \ .
\label{eq:ea}
\end{equation}
This quantity is a function of the particle densities $\rho_n,\rho_p,\rho_\Lambda$, and $\rho_{\Sigma^-}$ or,
equivalently, of the total baryonic density $\rho=\rho_n+\rho_p+\rho_\Lambda+\rho_{\Sigma^-}$, the hyperon fraction
$Y=(\rho_{\Sigma^-}+\rho_\Lambda)/\rho$, the isospin asymmetry $\beta=(\rho_n-\rho_p)/(\rho_n+\rho_p)$, and the asymmetry 
between the $\Lambda$ and $\Sigma^-$ hyperons $\alpha=(\rho_{\Sigma^-}-\rho_\Lambda)/(\rho_{\Sigma^-}+\rho_\Lambda)$. 

As referred before, Brueckner-type calculations are very time consuming since one has to solve a self-consistent set of coupled-channel
equations for different strangeness sectors. Therefore, from a practical point of view, it would be interesting and useful to characterize 
the dependence of the total energy per particle $E/A$ on the particle densities $\rho_n,\rho_p,\rho_\Lambda$, and $\rho_{\Sigma^-}$, or, 
alternatively on $\rho, Y, \beta$ and $\alpha$, in a simple analytical form. The free Fermi gas contribution, $T/A$, is already analytic, reading
\begin{eqnarray}
\frac{T}{A}=\sum_{i=n,p,\Lambda,\Sigma^-}\frac{3}{5}\frac{\hbar^2k_{F_i}^2}{2M_i}\frac{\rho_i}{\rho}
=\frac{3}{5}\frac{\hbar^2k_F^2}{2}\frac{1}{2}\left[
 \frac{1}{M_n}\left(1-Y\right)^{5/3}\left(1+\beta\right)^{5/3}
+\frac{1}{M_p}\left(1-Y\right)^{5/3}\left(1-\beta\right)^{5/3} \right. \nonumber \\
\left. +\frac{1}{M_{\Lambda}}Y^{5/3}\left(1-\alpha\right)^{5/3}
+\frac{1}{M_{\Sigma^-}}Y^{5/3}\left(1+\alpha\right)^{5/3}
\right] \ ,
\label{eq:ta}
\end{eqnarray}
where $k_{F_i}=(3\pi^2\rho_i)^{1/3}$ and we have defined $k_F \equiv (3\pi^2\rho/2)^{1/3}$. An idea of the possible terms appearing 
in the correlation energy contribution, $V/A$, can be obtained from the following phase space analysis of the single-particle potentials, 
similar to the ones perfomed in Ref.\ \cite{bombaci91} for isospin asymmetric matter and in Ref.\ \cite{vidana02} for 
spin-polarized isospin asymmetric matter. Replacing the matrix elements 
$\langle \vec{k}\vec{k'}|G(\omega)_{B_iB_j,B_iB_j}(\omega=E_{B_i}+E_{B_j})|\vec{k}\vec{k'} \rangle$ by an average
value, $g_{B_iB_j}(\vec{k},\rho,Y,\beta,\alpha)$, in the Fermi sphere with radius $k'\leq k_F^{B_j}$, and integrating over the corresponding Fermi sea, the 
single-particle potentials of the four baryon species under consideration can be written as
\begin{equation}
\begin{array}{l}
U_n(\vec{k}) \sim g_{nn}\rho_n+ g_{np}\rho_p+ g_{n\Lambda}\rho_{\Lambda}+ g_{n\Sigma^-}\rho_{\Sigma^-} \ , \\
U_p(\vec{k}) \sim g_{pn}\rho_n+ g_{pp}\rho_p+ g_{p\Lambda}\rho_{\Lambda}+ g_{p\Sigma^-}\rho_{\Sigma^-} \ , \\
U_{\Lambda}(\vec{k}) \sim g_{\Lambda n}\rho_n+ g_{\Lambda p}\rho_p \ , \\
U_{\Sigma^-}(\vec{k}) \sim g_{\Sigma^- n}\rho_n+ g_{\Sigma^- p}\rho_p \ . \\
\end{array}
\label{eq:spp2}
\end{equation}
For small values of the hyperon fraction, the isospin asymmetry, and the asymmetry between $\Lambda$'s and $\Sigma^-$'s, one can neglect
the dependence on $Y, \beta$ and $\alpha$ of the average $G$ matrices assuming $g_{B_iB_j}(\vec{k},\rho,Y,\beta,\alpha)\sim g_{B_iB_j}(\vec{k},\rho)$, and 
\begin{equation}
\begin{array}{l}
g_{nn} \approx g_{pp} \equiv g_1(\vec{k},\rho) \ , \\
g_{np} \approx g_{pn} \equiv g_2(\vec{k},\rho) \ ,  \\
g_{n\Lambda} \approx g_{\Lambda n} \approx g_{p\Lambda} \approx g_{\Lambda p} \equiv g_3(\vec{k},\rho) \ , \\
g_{n\Sigma^-} \approx g_{\Sigma^-n} \equiv g_4(\vec{k},\rho) \ , \\
g_{p\Sigma^-} \approx g_{\Sigma^-p} \equiv g_5(\vec{k},\rho) \ . 
\end{array}
\label{eq:gi}
\end{equation}
Note that the quantities $g_i(\vec{k},\rho)$ receive contributions from different isospin ($T$) and strangeness ($S$) channels. Whereas
$g_1(\vec{k},\rho)$ receives contributions only from the isospin triplet and zero strangeness channel, $g_2(\vec{k},\rho)$ has in addition
a contribution from the isospin singlet, $g_3(\vec{k},\rho)$ and $g_4(\vec{k},\rho)$ are, respectively, purely isospin $1/2$ and $3/2$ with 
strangeness $-1$, and $g_5(\vec{k},\rho)$ has contributions from $T=1/2$ and $T=3/2$ with $S=-1$. 

\begin{figure*}[h]
\begin{center}
\includegraphics[width=12.cm]{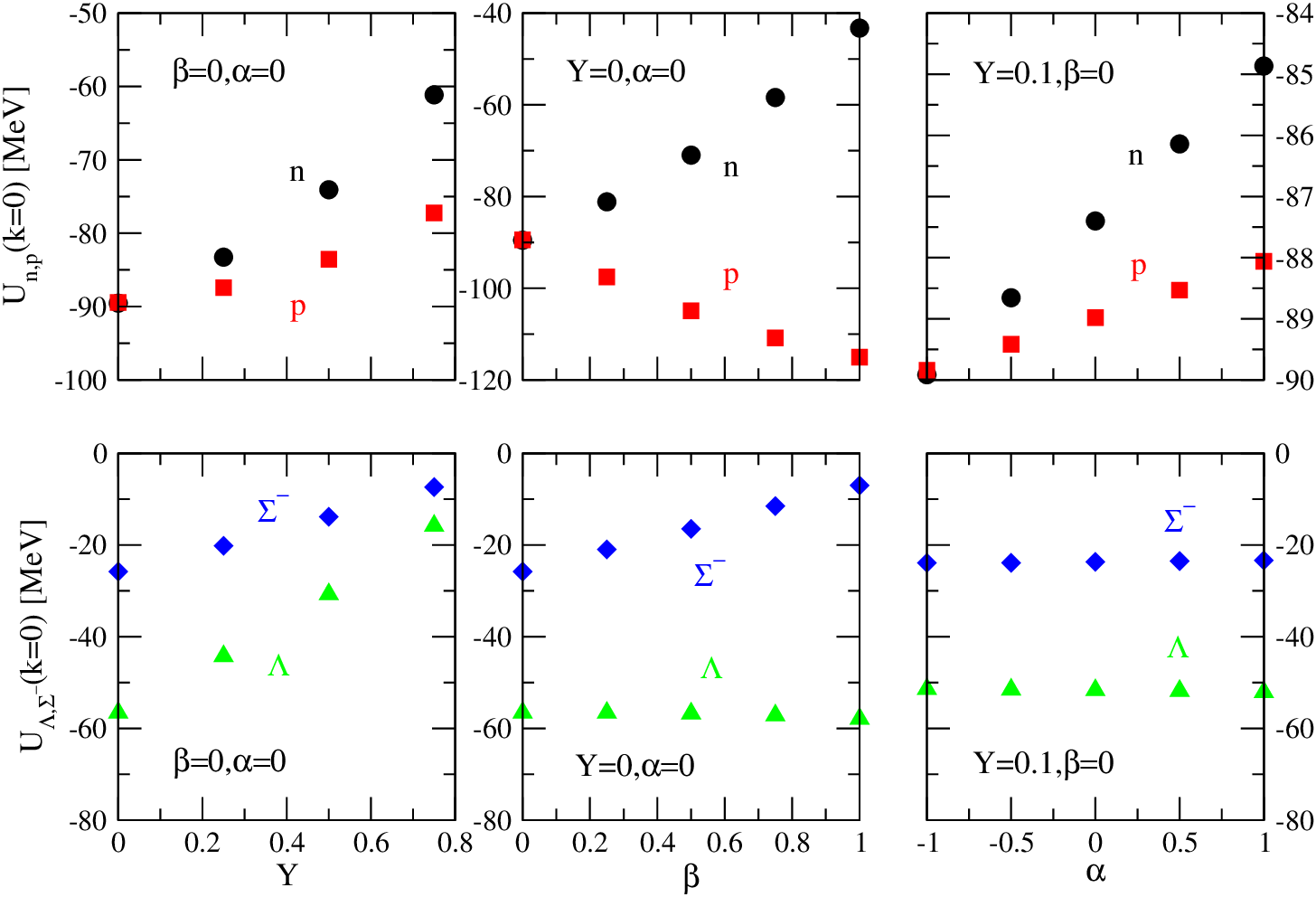}
\caption{(Color online) Neutron (circles), proton (squares), $\Lambda$ (triangles) and $\Sigma^-$ (diamonds)
single-particle potentials at $\vec{k}=0$ and $\rho=\rho_0$ as a function of $Y$ (left panels), $\beta$ (middle panels) and $\alpha$ 
(right panels) obtained with the J\"{u}lich model. Upper (lower) panels show the results for neutrons and protons ($\Lambda$ and $\Sigma^-$).}
\label{fig:spp}
\end{center}
\end{figure*}

Using the set of Eqs.\ (\ref{eq:spp2}) and (\ref{eq:gi}), the single-particle potentials can then be written as
\begin{equation}
U_n(\vec{k}) \sim \frac{\rho}{2}\left(1-Y\right)\left[g_1(\vec{k},\rho)\left(1+\beta\right) + g_2(\vec{k},\rho)\left(1-\beta\right)\right]
                      + \frac{\rho}{2}Y\left[g_3(\vec{k},\rho)\left(1-\alpha\right)+ g_4(\vec{k},\rho)\left(1+\alpha\right)\right] \ , 
\label{eq:spn}
\end{equation}
\begin{equation}
U_p(\vec{k}) \sim \frac{\rho}{2}\left(1-Y\right)\left[g_2(\vec{k},\rho)\left(1+\beta\right) + g_1(\vec{k},\rho)\left(1-\beta\right)\right]
                      + \frac{\rho}{2}Y\left[g_3(\vec{k},\rho)\left(1-\alpha\right)+ g_5(\vec{k},\rho)\left(1+\alpha\right)\right] \ , 
\label{eq:sppt}
\end{equation}
\begin{equation}
U_{\Lambda}(\vec{k}) \sim \rho g_3(\vec{k},\rho)\left(1-Y\right) \ , 
\label{eq:spl}
\end{equation}
\begin{equation}
U_{\Sigma^-}(\vec{k}) \sim \frac{\rho}{2}\left(1-Y\right)\left[g_4(\vec{k},\rho)\left(1+\beta\right)+g_5(\vec{k},\rho)\left(1-\beta\right)\right] \,
\label{eq:spsm}
\end{equation}
where the particle densities $\rho_n,\, \rho_p,\, \rho_{\Lambda}$ and $\rho_{\Sigma^-}$ have been written in terms of $\rho,\, \beta,\, Y$ and $\alpha$. 
These equations show explicitely the dependence of the single-particle potentials on the hyperon fraction, the isospin asymmetry and the asymmetry
between $\Lambda$'s and $\Sigma^-$'s. This dependence is tested in Fig.\ \ref{fig:spp} for the J\"{u}lich model where the value at $\vec{k}=0$ of the single-particle potentials 
$U_n, U_p, U_{\Lambda}$ and $U_{\Sigma^-}$ at the saturation density ($\rho_0=0.175$ fm$^{-3}$ in our model) is plotted as a function of $Y$ (with $\beta=\alpha=0$), 
$\beta$ (with $Y=\alpha=0$) and $\alpha$ (with $Y=0.1, \beta=0$), in the left, middle and right panels, respectively. Similar dependences has been obtained
also for the NSC89 and NSC97 models. The above equations predict a linear variation of the single-particle potentials with respect to  $Y, \beta $ and $\alpha$. 
This prediction is quite well confirmed from the microscopic results reported in Fig. \ref{fig:spp}, although deviations from the linear behavior are found at 
higher values of $Y, \beta $ and $\alpha$. These deviations have to be associated to the dependence of the average $G$ matrices on $Y, \beta $ and $\alpha$, which 
has been neglected in the present analysis (see the set of Eqs.\ (\ref{eq:gi})). 

Now, using Eqs.\ (\ref{eq:spn})-(\ref{eq:spsm}), and replacing the quantities $g_i(\vec{k},\rho)$ by their averages ${\bar g}_i(\rho)$ in the corresponding Fermi spheres, 
one can see, after integration, that the correlation energy behaves like
\begin{eqnarray}
\frac{V}{A} \sim  
  \frac{{\bar g}_1(\rho)}{2\rho}\left(\rho_n^2+\rho_p^2\right)
+\frac{{\bar g}_2(\rho)}{\rho}\rho_n\rho_p
+\frac{{\bar g}_3(\rho)}{\rho}\left(\rho_n+\rho_p\right)\rho_{\Lambda}
+\frac{{\bar g}_4(\rho)}{\rho}\rho_n\rho_{\Sigma^-}
+\frac{{\bar g}_5(\rho)}{\rho}\rho_p\rho_{\Sigma^-} 
\ ,
\label{eq:va}
\end{eqnarray}
or, replacing the particle densities in terms of $\rho, \beta, Y$ and $\alpha$
\begin{eqnarray}
\frac{V}{A} \sim \frac{\rho}{4}{\bar g_1}(\rho)\left(1-Y\right)^2\left(1+\beta^2\right)
           +\frac{\rho}{4}{\bar g}_2(\rho)\left(1-Y\right)^2\left(1-\beta^2\right)
           +\frac{\rho}{2}{\bar g}_3(\rho)Y\left(1-Y\right)\left(1-\alpha\right) \nonumber \\
           +\frac{\rho}{4}{\bar g}_4(\rho)Y\left(1-Y\right)\left(1+\alpha\right)\left(1+\beta\right)
           +\frac{\rho}{4}{\bar g}_5(\rho)Y\left(1-Y\right)\left(1+\alpha\right)\left(1-\beta\right) \ .
\label{eq:va2}
\end{eqnarray}

\begin{figure*}[h]
\begin{center}
\includegraphics[width=12.cm]{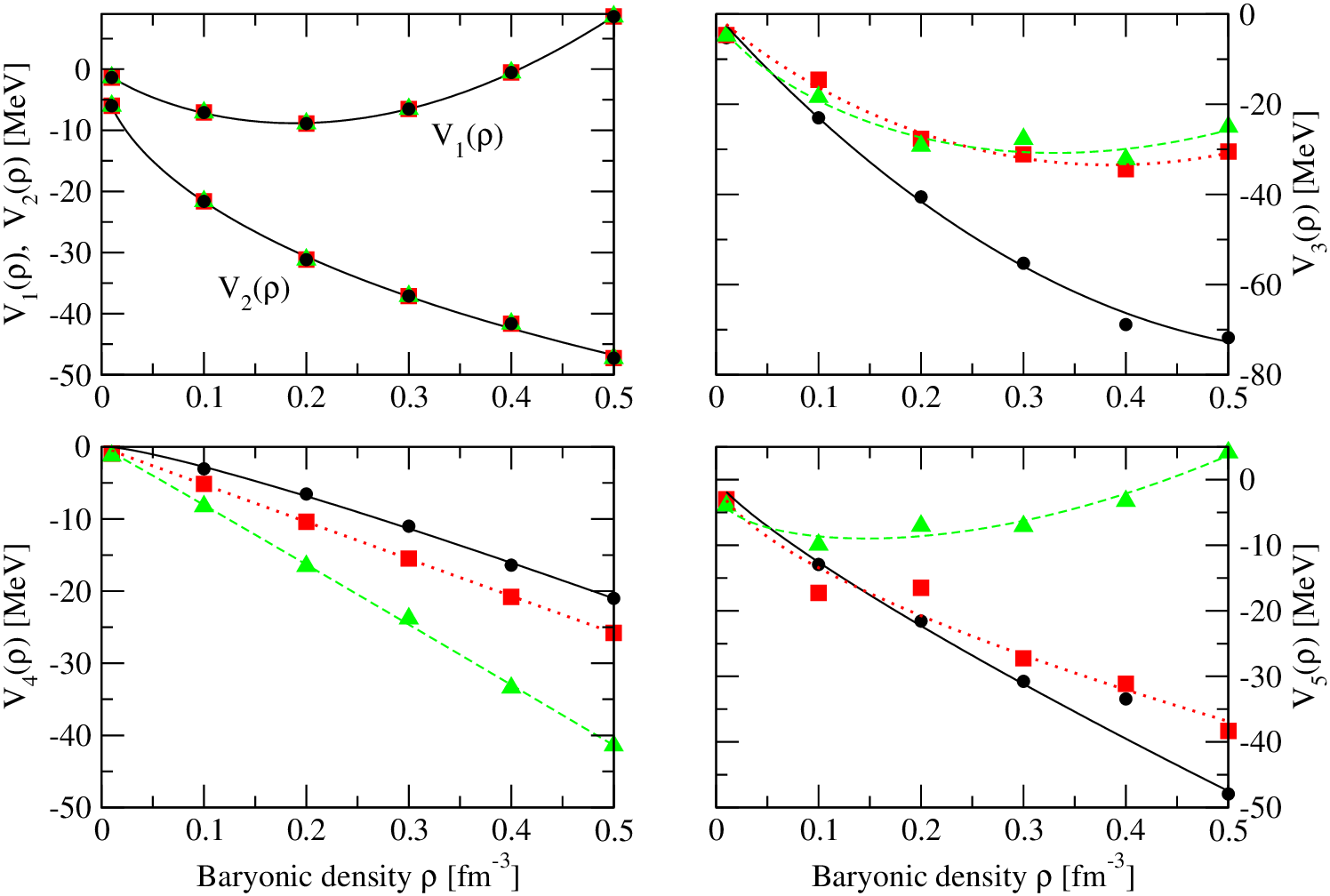}
\caption{(Color online) Density dependence of the coefficients $V_i(\rho)$
of Eq.\ (\ref{eq:ea2}). Circles, squares and triangles show the result of the microscopic BHF calculation obtained
with the J\"{u}lich, NSC89 and NSC97e models, respectively, whereas solid, dotted and dashed lines refer to the 
parametrization defined in Eq.\ (\ref{eq:coef}) and Tables \ref{tab1}--\ref{tab3}.}
\label{fig:para}
\end{center}
\end{figure*}

\begin{figure*}[h]
\begin{center}
\includegraphics[width=12.cm]{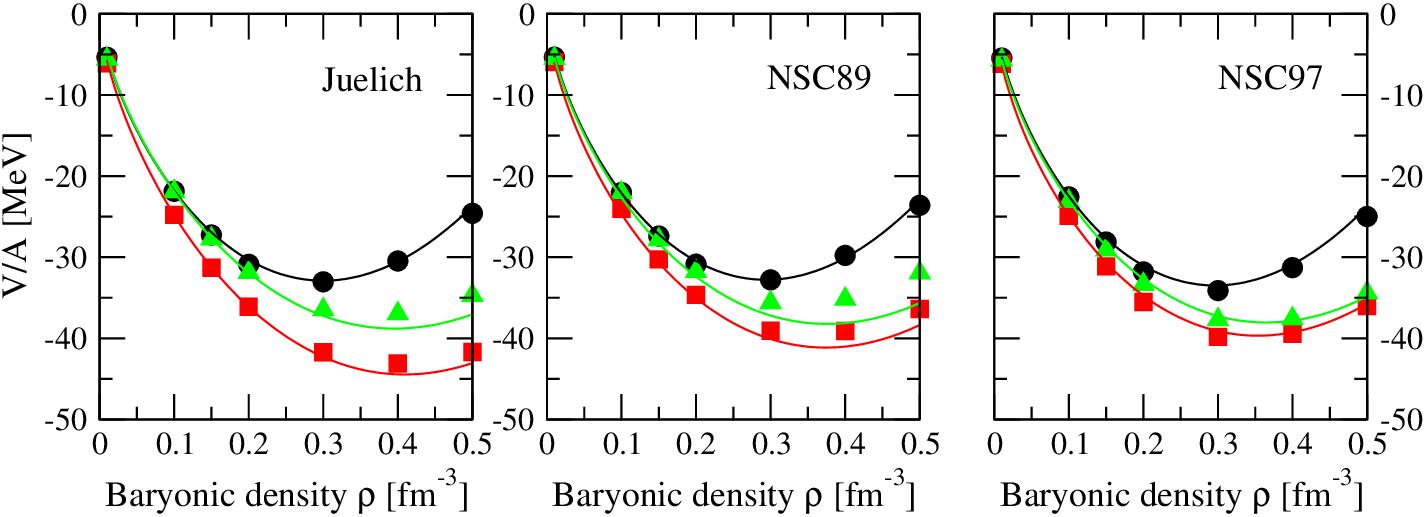}
\caption{(Color online) Correlation energy $V/A$ as a function of the densty for three
arbitrary sets of values of $Y, \beta$ and $\alpha$: $Y=0.08, \beta=0.6, \alpha=0.6$ (circles),
$Y=0.15, \beta=0.2, \alpha=0.2$ (squares) and $Y=0.14, \beta=0.4, \alpha=0.75$ (triangles). 
Symbols show the result of the microscopic BHF calculation, whereas dashed lines refer to the 
parametrization defined according to Eqs.\ (\ref{eq:ea2}) and (\ref{eq:coef}), and Tables
\ref{tab1}--\ref{tab3}.}
\label{fig:va2}
\end{center}
\end{figure*}

From this simple analysis we can finally infer the form of the correlation energy
\begin{eqnarray}
\frac{V}{A}=V_1(\rho)\left(1-Y\right)^2\left(1+\beta^2\right)
           +V_2(\rho)\left(1-Y\right)^2\left(1-\beta^2\right)
           +V_3(\rho)Y\left(1-Y\right)\left(1-\alpha\right) \nonumber \\
           +V_4(\rho)Y\left(1-Y\right)\left(1+\alpha\right)\left(1+\beta\right)
           +V_5(\rho)Y\left(1-Y\right)\left(1+\alpha\right)\left(1-\beta\right) \ .
\label{eq:ea2}
\end{eqnarray}
The coefficients $V_1(\rho), V_2(\rho), V_3(\rho), V_4(\rho)$ and $V_5(\rho)$ have been fitted to reproduce the microscopic BHF results
corresponding to the following five set of values of $Y, \beta$ and $\alpha$: ($Y=0,\beta=0,\alpha=0$), ($Y=0,\beta=1,\alpha=0$), ($Y=0.1,\beta=0.875,\alpha=1$),
($Y=0.15,\beta=0.7,\alpha=0.5$), and ($Y=0.2,\beta=0.5,\alpha=0$). The first two sets guarantee that the parametrization reproduces the microscopic results for 
symmetric nuclear matter and pure neutron matter. The other three have been chosen in order to mimic three representative $\beta$-stable matter compositions 
for densities above the hyperon threshold obtained with microscopic approaches (see {\it e.g.,} Refs.\ \cite{baldo00} and \cite{vidana00b}). It is clear that the 
determination of these coefficients is not unique. However, we have checked that with the choice of this set of values of $Y, \beta$ and $\alpha$, we
get a parametrization that reproduces with good quality (see Fig.\ \ref{fig:va2} and the discussion below) the results of the BHF calculation for a
wide range of arbitrary values of $Y, \beta$ and $\alpha$. In addition, we have adjusted the density dependence of the coefficients $V_i(\rho)$ in the 
following functional form
\begin{equation}
V_i(\rho)=a\rho^\gamma+b\rho^\delta \,\,\,\, i=1-5 \ ,
\label{eq:coef}
\end{equation}
where the set of parameters $a, \gamma, b$ and $\delta$ for the five coefficients are given in Tables \ref{tab1}--\ref{tab3}. The density dependence of the coefficients $V_i(\rho)$ together 
with the functional defined in the above  equation is shown in Fig.\ \ref{fig:para} for the three models considered. It is seen that the functional of Eq.\ (\ref{eq:coef}) 
reproduces reasonably well the microscopic results for the three models in the whole range of densities explored (0.01 fm$^{-3}$ $ < \rho <$ 0.5 fm$^{-3}$).

\begin{figure*}[h]
\begin{center}
\includegraphics[width=12.cm]{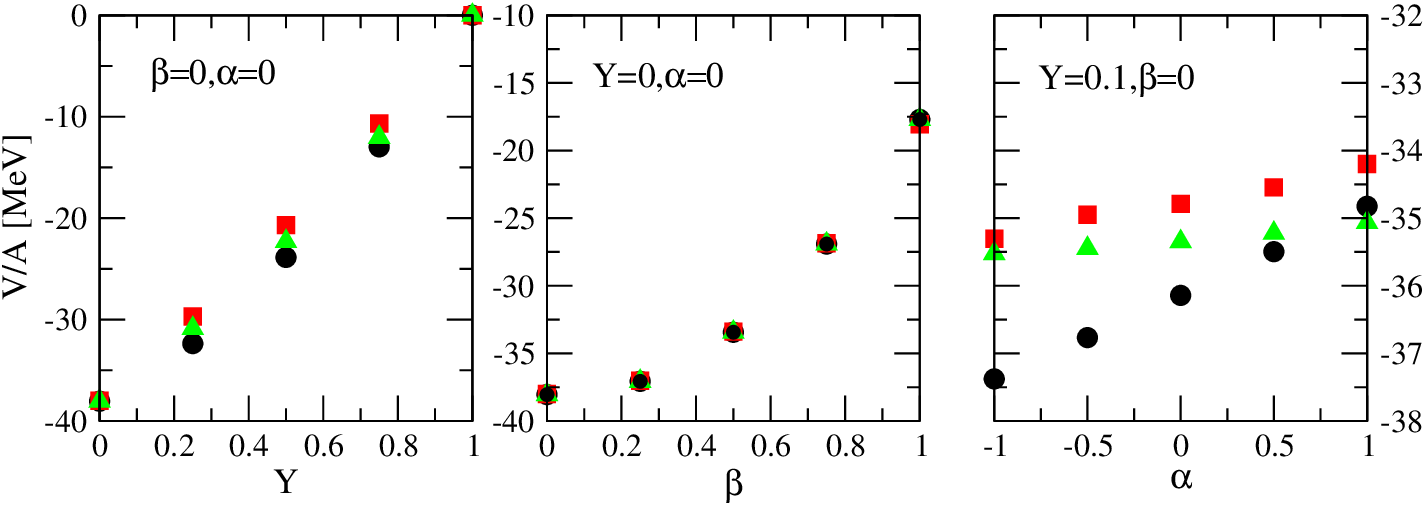}
\caption{(Color online) Correlation energy $V/A$  at $\rho=\rho_0$ as a function of $Y$ (left panel), $\beta$ (middle panel) 
and $\alpha$ (right panel). Circles, squares and triangles show the result of the microscopic BHF calculation obtained
with the J\"{u}lich, NSC89 and NSC97e models, respectively.}
\label{fig:va}
\end{center}
\end{figure*}

In order to test the quality of our parametrization, we show for the three models, in the left, middle and right panels of Fig.\ \ref{fig:va}, the correlation energy 
at $\rho=\rho_0$ as a function of $Y$ (with $\beta=\alpha=0$), $\beta$ (with $Y=\alpha=0$) and $\alpha$ (with $Y=0.1, \beta=0$). Circles, squares and triangles show 
the result of the microscopic BHF calculation obtained with the J\"{u}lich, NSC89 and NSC97e models, respectively. From the figure it can be seen that the dependence 
on $Y, \beta$ and $\alpha$ predicted by Eq.\ (\ref{eq:ea2}) (quadratic in $Y$ and $\beta$, and linear in $\alpha$) is well confirmed from the microscopic results. 
For completeness, we finally compare in Fig.\ \ref{fig:va2}, the results for the correlation energy as a function of the density for three arbitrary sets of values 
of $Y, \beta$ and $\alpha$: (i) $Y=0.08,\beta=0.6,\alpha=0.6$, (ii) $Y=0.15,\beta=0.2,\alpha=0.2$, and (iii) $Y=0.17,\beta=0.4,\alpha=0.75$, obtained from the microscopic 
BHF calculation (symbols) and from the parametrization (lines) for the three models considered. The quality of the parametrization is quite good for the three 
models, as it can be seen in the figure, with deviations from the microscopic calculation, at the higher density explored, of at most $2\%$ and $5-6\%$ for the NSC97 
and J\"{u}lich models, respectively, and of about $10-11\%$ for the NSC89 one. These deviations, as it has been said before, have to be associated to the dependences on 
$Y, \beta$ and $\alpha$ neglected in the construction of the parametrization.

To summarize, we have constructed an analytic parametrization of the hyperonic matter equation of state based on microscopic Brueckner--Hartree--Fock calculations using the 
realistic Argonne V18 NN potential plus a NNN TBF of Urbana type, and three models of the YN interaction: the Nijmegen soft-core models NSC89 and
NSC97e, and the most recent meson-exchange potential of the J\"{u}lich group. The construction of this parametrization is based on a simple phase-space 
analysis, and reproduces with good accuracy the results of the microscopic calculations, allowing for rapid calculations that accurately mimic the microscopic BHF results, being, thus, 
very useful from a practical point of view. Our parametrization will be extended in a future work to include finite temperature effects, necessary to describe the properties of newborn 
neutron stars, and the conditions of matter in relativistic heavy-ion collisions.


\section*{Acknowledgments}

We are very grateful to \`Angels Ramos for useful former discussions. This work has been partially supported by FCT (Portugal) 
under grants SFRH/BD/62353/2009 and FCOMP-01-0124-FEDER-008393 with FCT reference CERN/FP/109316/2009, the Consolider 
Ingenio 2010 Programme CPAN CSD2007-00042 and Grant No. FIS2008-01661 from MEC and FEDER (Spain) and Grant 2009GR-1289
from Generalitat de Catalunya (Spain), and by COMPSTAR, an ESF (European Science Foundation) Research Networking 
Programme.


\newpage

\newpage

\newpage

\newpage

\newpage

\begin{center}
\begin{table*}
\begin{tabular}{c|cccc}
\hline
\hline
Coefficient  & $a$ & $\gamma$ & $b$ & $\delta$ \\
\hline
$V_1$ &  $-65.0189$ & $0.843983$ &  $166.944$ &  $1.89579$ \\
$V_2$ &  $-144.122$ & $0.628802$ &  $82.4707$ &  $0.829031$ \\
$V_3$ &  $-241.211$ & $0.984562$ &  $195.95$  &  $1.99311$ \\
$V_4$ &  $-123.882$ & $0.999992$ &  $76.3707$ &  $0.90001$ \\
$V_5$ &  $-98.8994$ & $0.82631$9 &  $14.7032$ &  $0.830927$ \\
\hline
\hline
\end{tabular}
\caption{Set of parameters $a, \gamma, b$ and $\delta$ characterizing the density dependence of the coefficients $V_i(\rho)$ for the J\"{u}lich model. The parameters
$\gamma$ and $\delta$ are dimensionless, whereas the units of $a$ and $b$ are MeV $\times$ fm$^{3\gamma}$ and  MeV $\times$ fm$^{3\delta}$, respectively.}
\label{tab1}
\end{table*}
\end{center}

\newpage

\begin{center}
\begin{table*}
\begin{tabular}{c|cccc}
\hline
\hline
Coefficient & $a$ & $\gamma$ & $b$ & $\delta$ \\
\hline
$V_1$ &  $-65.0189$ & $0.843983$ &  $166.944$ &  $1.89579$ \\
$V_2$ &  $-144.122$ & $0.628802$ &  $82.4707$ &  $0.829031$ \\
$V_3$ &  $-142.908$ & $0.897287$ &  $183.341$ &  $1.99791$ \\
$V_4$ &  $-123.583$ & $0.987746$ &  $72.1931$ &  $0.983861$ \\
$V_5$ &  $-114.735$ & $0.629932$ &  $57.809$  &  $0.632833$ \\
\hline
\hline
\end{tabular}
\caption{Same as Table \ref{tab1} for the NSC89 model.}
\label{tab2}
\end{table*}
\end{center}

\newpage

\begin{center}
\begin{table*}
\begin{tabular}{c|cccc}
\hline
\hline
Coefficient & $a$ & $\gamma$ & $b$ & $\delta$ \\
\hline
$V_1$ &  $-65.0189$ & $0.843983$ &  $166.944$ &  $1.89579$ \\
$V_2$ &  $-144.122$ & $0.628802$ &  $82.4707$ &  $0.829031$ \\
$V_3$ &  $-108.951$ & $0.71094$  &  $149.79$  &  $1.8744$ \\
$V_4$ &  $-95.221$  & $0.999994$ &  $11.6305$ &  $0.901149$ \\
$V_5$ &  $-21.5452$ & $0.34935$  &  $76.0311$ & $1.88421$ \\
\hline
\hline
\end{tabular}
\caption{Same as Table \ref{tab1} for the NSC97e model.}
\label{tab3}
\end{table*}
\end{center}


\end{document}